\documentclass[pra,onecolumn,amsmath, superscriptaddress,notitlepage,showpacs,11pt]{revtex4-1}
\usepackage{amsmath,amsfonts,amssymb,amsthm,amstext,amscd,amsthm,dsfont,bbm,hyperref,natbib}
\usepackage{physics}
\usepackage{color}
\usepackage{soul,xcolor}
\usepackage{gensymb}
\usepackage{scalerel}
\hypersetup{
    colorlinks = true,
    citecolor=blue,
    linkcolor = red,
    anchorcolor = red,
    citecolor = blue,
    filecolor = red,	
    pagecolor = red,
    urlcolor = blue,
    }
\usepackage[normalem]{ulem}
\usepackage{graphicx}
\usepackage{bbold}
\usepackage{braket}
\usepackage{placeins}

\begin{document}

\title{Dynamics of quantum correlations in a Qubit-Oscillator system interacting via a dissipative bath}
	\author{Revanth Badveli}
	\affiliation{
		Computer Science and Information Systems, BITS Pilani-Goa Campus, Goa 403 726, India}
		\affiliation{Quantum Research Group, School  of Chemistry and Physics,
		University of KwaZulu-Natal, Durban 4001, South Africa}
	\author{Vinayak Jagadish}
	\affiliation{Quantum Research Group, School  of Chemistry and Physics,
		University of KwaZulu-Natal, Durban 4001, South Africa}\affiliation{ National
		Institute  for Theoretical  Physics  (NITheP), KwaZulu-Natal,  South
		Africa}
	\author{S.~Akshaya}
	\affiliation{Department of Physics, Lovely Professional University-Punjab- 144411, India}
			\author{R. Srikanth}
	\affiliation{Poornaprajna Institute of Scientific Research,
		Bangalore- 560 080, India}
	\author{Francesco Petruccione}
	\affiliation{Quantum Research Group, School  of Chemistry and Physics,
		University of KwaZulu-Natal, Durban 4001, South Africa}\affiliation{ National
		Institute  for Theoretical  Physics  (NITheP), KwaZulu-Natal,  South
		Africa}
\date{} 

\begin{abstract} 
The entanglement dynamics in a bipartite system consisting of a qubit and a harmonic oscillator interacting
only through their coupling with the same bath is studied. The considered model assumes that the qubit is coupled
to the bath via the Jaynes-Cummings interaction, whilst the position of the oscillator is coupled to the position of
the bath via a dipole interaction. We give a microscopic derivation of the Gorini-Kossakowski-Sudarshan-Lindblad equation for the considered model. Based on the Kossakowski Matrix, we show that non-classical correlations including entanglement can be generated by the considered dynamics.  We then analytically identify specific initial states for which entanglement is generated. This result is also supported by our numerical simulations.
\end{abstract}
\maketitle  
\section{Introduction}
\setcounter{equation}{0}

The interactions with the environment are the fundamental source of noise in quantum systems that can thwart attempts to exploit intrinsic quantum properties-entanglement and coherence-for quantum computing and communication and hence the study of open quantum systems \cite{petruccione,rivas2012open,haroche_exploring_2006,Quanta77} is an active area of research.
But in some scenarios \cite{ma2012entanglement,braun2002creation,contreras2008entanglement,banerjee2010dynamics,banerjee2010entanglement,paz2008dynamics,su2014scheme,utami2008entanglement}, the environment can mediate entanglement generation. Previously investigated examples of such scenarios include two qubits interacting with a Lorentz-broadened cavity mode at zero temperature \cite{ma2012entanglement}, two qubits interacting with infinitely many degrees of freedom of common heat bath in thermal equilibrium \cite{braun2002creation}, and two charge qubits strongly coupled to common boson bath \cite{contreras2008entanglement}. In \cite{ma2012entanglement}, both the Born and Rotating Wave Approximations are avoided. In \cite{contreras2008entanglement}, Markovian approximations for coupling to electronic reservoirs and non-Markovian approximations for strong coupling with boson bath is used. In  \cite{banerjee2010dynamics,banerjee2010entanglement}, the entanglement dynamics of two-qubit system interacting with a squeezed thermal bath via dissipative interaction and non-demolition interaction is studied, respectively. In \cite{paz2008dynamics}, the evolution of entanglement between two resonant oscillators is characterized and the phases of sudden death and revival of entanglement are identified. In \cite{su2014scheme}, a dissipative scheme, which is independent of initial states, is proposed to generate maximal entanglement between two atoms trapped in an optical cavity. In \cite{utami2008entanglement}, an analytical expression of logarithmic negativity is derived for a system  of a qubit dispersively coupled to a dissipative oscillator.

The coupling between a two-level system and a harmonic oscillator has been shown to give rise to many interesting effects in ion trap \cite{leibfried2003quantum,cirac1995quantum,wineland2003quantum,sorensen1999quantum} and Cavity Quantum Electrodynamics experiments \cite{raimond2001manipulating,osnaghi2001coherent}. The two-level system can be used to generate and probe non-classical states of the oscillator. Reciprocally,
the oscillator has been used as a \textquotedblleft
catalyst\textquotedblright\ to produce entanglement between multiple
qubits. Generating entanglement between qubits without any direct interaction can have applications in Quantum Cryptography \cite{cubitt2003separable}.

In this paper we focus on a bipartite system consisting of a two level system (qubit) and a harmonic oscillator interacting with the same bath of harmonic oscillators. The qubit is interacting with the bath via Jaynes-Cummings interaction type, whereas the position of the oscillator is coupled to the position of the bath via dipole interaction. The schematic diagram for the considered model is shown in Figure \ref{fig13}. A microscopic derivation is given for the master equation describing the open dynamics of the system considered in this model, which turns out to be in the form of the Gorini-Kossakowski-Sudarshan-Lindblad (GKSL) equation \cite{gorini1976completely,lindblad1976generators}. We follow \cite{hu2011necessary,benatti2008environment,benatti2003environment} to show that this bath mediated interaction can lead to the generation of non-classical correlations including entanglement. We then find out the class of initial states which get entangled due to this bath mediated interaction.
The dependence of entanglement generation on temperature of the bath in this scenario is shown. The entanglement measure, Negativity \cite{eisert2006entanglement,vidal2002computable} is used to study the entanglement dynamics of our system with respect to the relative coupling strength of qubit-bath coupling and oscillator-bath coupling. Weaker forms of quantum correlations are also studied using mutual information and quantum discord.

The outline of the paper is as follows. In Sect.~2 we express the Hamiltonian of the system of qubit and the  harmonic oscillator and its interaction Hamiltonian. We then derive the master equation for the system. In Sect.~3 we discuss the possibility of generation of quantum correlations in the system and we find out the classes of initial which lead to entanglement. In Sect.~4 we study the dynamics of entanglement, mutual information and discord. We conclude in Sect.~5.  
\section{Microscopic Derivation of the Master Equation}
We work in units of $\hbar$ and $k_B$ set to 1. The total Hamiltonian $H_T$
for the composite system is, 
\begin{equation}
H_T = H_{S}+H_{B}+H_{I},
\end{equation}
where $H_{S}$, $H_{B}$ and $H_{I}$ are the Hamiltonian for the
system, bath and interaction respectively. The system under consideration is a qubit and the harmonic oscillator interacting resonantly with the bath. The system Hamiltonian is therefore a sum of $H_{Q}$ and $ H_{HO}$, that of the qubit and the harmonic oscillator, respectively. 
\begin{align}
H_{S} &= H_{Q} + H_{HO}.
\end{align} Introducing ladder operators $a$ and $a^{\dagger}$ for the oscillator and excluding the zero-point energy, $H_{S}$ can be written as
\begin{equation}
H_S = \frac{\Omega}{2}\sigma_{z} +\Omega a^{\dagger}a.
\end{equation} 
The bath is modeled as a collection of independent harmonic oscillators and the Hamiltonian is \begin{equation}
H_{B} = \sum_{k}
\omega_{k} b_{k}^{\dagger} b_{k},
\end{equation} where  $b_{k}^{\dagger}$ and $b_{k}$ are the bosonic creation and annihilation operators obeying the standard commutation relation $[b_{k}, b_{k'}^{\dagger}] = \delta_{k,k'}$. The interaction between the system and the bath is assumed as follows. The qubit interacts with the bath via a Jaynes-Cummings interaction type, which indicates a radiative damping by its interaction with the many modes of the bath of harmonic oscillators. Also, the position of the oscillator is coupled to the position of the bath via dipole interaction. After the rotating-wave approximations are introduced, the total interaction Hamiltonian is expressed as
\begin{equation}
H_{I} = \sum_{k} g_{1} (\sigma_{+}b_{k} + \sigma_{-}{b}_{k}^{\dagger}) + g_{2} (a^{\dagger} b_{k} + a b_{k}^{\dagger}),
\end{equation} 
where $\sigma_{+(-)}$ is the qubit raising (lowering) operator. $g_{1 }$ and  $g_{2 }$ are the coupling constants corresponding to qubit-bath coupling and oscillator-bath coupling respectively. The interaction Hamiltonian in the interaction picture, $\tilde{H}_{I}$ is obtained as  
\begin{align}
\tilde{H}_{I} = \sum_{k}
g_{1} (\sigma_{+} e^{\imath\Omega t}b_{k}e^{-\imath\omega_{k} t} + \sigma_{-}e^{-\imath\Omega t}{b}_{k}^{\dagger}e^{\imath\omega_{k} t}) \nonumber\\+ g_{2}(a^{\dagger}e^{\imath\Omega t}b_{k}e^{-\imath\omega_{k} t} + ae^{-\imath\Omega t} b_{k}^{\dagger}e^{\imath\omega_{k} t}).
\end{align}
Introducing the operators\\
\begin{figure*}
	\begin{center}
		\includegraphics[scale = 0.6]{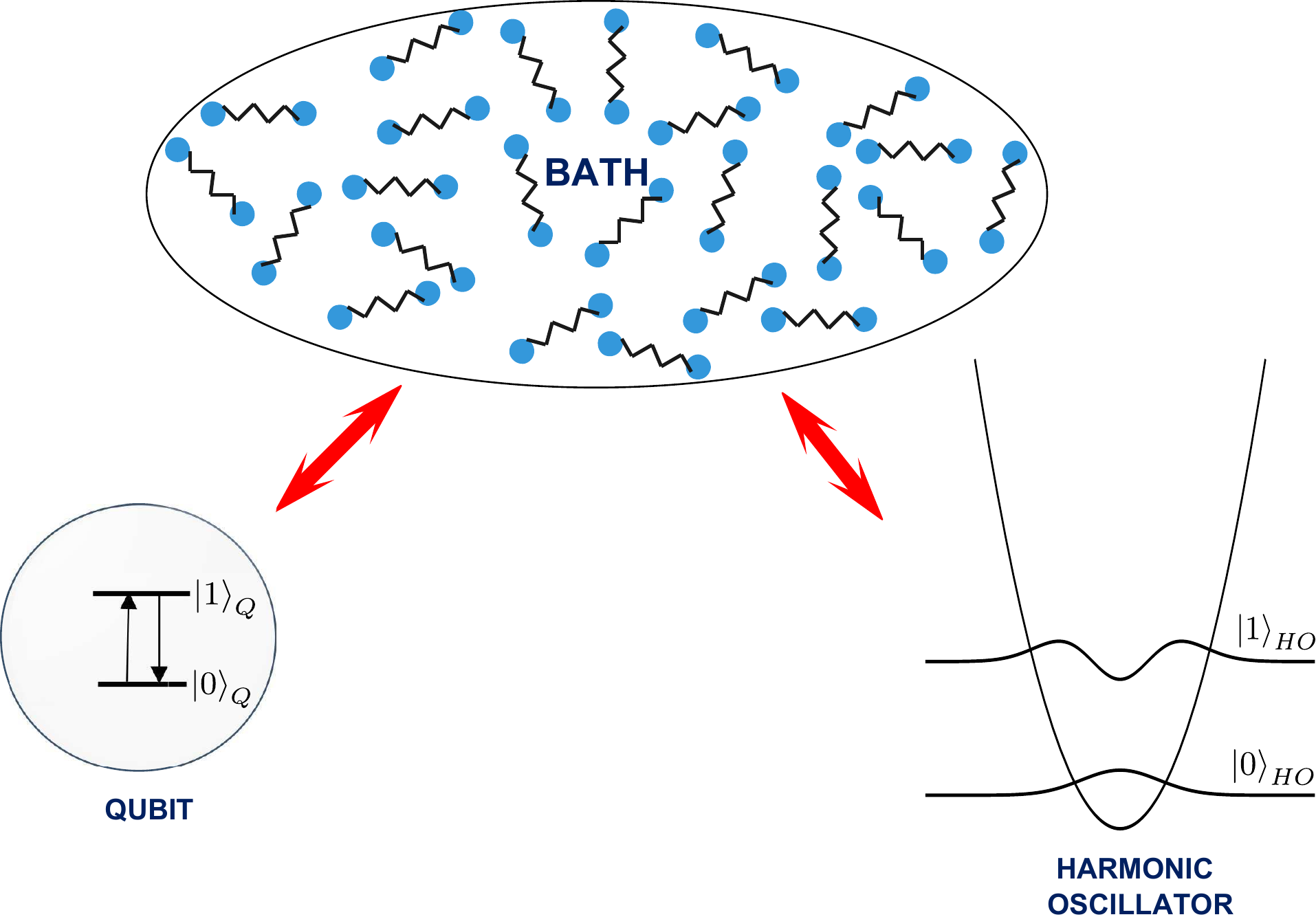}\hspace{5pt}
		\caption{(Color online) The schematic diagram of the considered model. The system consists of a two level system (qubit) and a harmonic oscillator, both of which interact with a bath of harmonic oscillators.} \label{fig13}
	\end{center}
\end{figure*}
\begin{eqnarray}
\label{eq:S_i operators}
S_{1} = \sigma_{+}  e^{\imath\Omega t},\thinspace \thinspace \thinspace&&   B_{1} = \sum_{k} g_{1}b_{k}e^{-\imath\omega_{k} t}, \nonumber\\
S_{2} = \sigma_{-}  e^{-\imath\Omega t}, \thinspace \thinspace \thinspace&&   B_{2} = \sum_{k} g_{1}b_{k}^{\dagger}e^{\imath\omega_{k} t}, \nonumber\\
S_{3} = a^{\dagger}  e^{\imath\Omega t},\thinspace \thinspace \thinspace&&  B_{3} = \sum_{k} g_{2}b_{k}e^{-\imath\omega_{k} t}, \nonumber\\
S_{4} =  a  e^{-\imath\Omega t},\thinspace \thinspace \thinspace&&   B_{4} = \sum_{k} g_{2}b_{k}^{\dagger}e^{\imath\omega_{k} t}. \nonumber\\
\end{eqnarray}
The interaction Halmiltonian $\tilde{H}_{I}$ can be written as
\begin{eqnarray}
\label{eq:interactiondecomp}
\tilde{H}_{I} = \sum_{i=1}^4 S_{i} B_{i},
\end{eqnarray} where $S$ and $B$ refer to operators pertaining to the space of operators on the system and bath respectively. Assuming weak coupling with the bath, we use the Born-Markov approximation to obtain the master equation for the reduced density matrix of the system.

Defining the bath-correlation functions as $F_{ij} (t,s) = \langle B_{i}(t)B_{j}(s)\rangle = \mbox{Tr}_{B} (B_{i}(t)B_{j}(s)\rho_B)$ and assuming that the bath is in an equilibrium state so that $F_{ij} (t,s) = F_{ij} (t-s)$ and using the form of Eq.~(\ref{eq:interactiondecomp}), one obtains the master equation in the Schr\"{o}dinger picture as follows.  

\begin{eqnarray}
\label{eq:twooneone}
\frac{\partial\rho_S(t)}{\partial t}=-\imath~
[H_S,\rho_S(t)] -\sum_{\substack{i, j = 1 \\}}^{4}\int_0^{\infty}d\tau~\Big([S_{i},S_{j}(-\tau)\rho_S(t)]F_{ij}(\tau)\nonumber\\+ [\rho_S(t)S_{j}(-\tau),S_{i}]F_{ji}(-\tau) \Big).
\end{eqnarray}
Let's define 
\begin{eqnarray}
\int_0^{\infty}d\tau e^{\imath\Omega \tau}F_{12}(\tau) &= \Gamma_{1} = \gamma_{1} + \imath\delta_1,\\
\int_0^{\infty}d\tau e^{\imath\Omega \tau}F_{21}(-\tau) &= \Gamma_{2} = \gamma_{2} + \imath\delta_2.
\end{eqnarray}
where $\{ \gamma_{1},\gamma_{2},\delta_1,\delta_2\} \in \mathbbm{R}$. The real parts $\gamma_1$ and $\gamma_{2}$ are given by $\zeta\frac{\exp(1/T)}{\exp(1/T) - 1} $ and  $ \zeta\frac{1}{\exp(1/T) - 1} $ respectively, where $\zeta$ is the spontaneous emission constant and temperature of the bath $T$ is given in the units of $\Omega$.
Now one can see that 
\begin{eqnarray}
\label{eq:Fij}
\int_0^{\infty}d\tau e^{\imath\Omega \tau}F_{41}(-\tau) = \int_0^{\infty}d\tau e^{\imath\Omega \tau}F_{23}(-\tau) = \frac{g_{2}}{g_{1}}\Gamma_{2} ,\\ 
\int_0^{\infty}d\tau e^{\imath\Omega \tau}F_{14}(\tau) = \int_0^{\infty}d\tau e^{\imath\Omega \tau}F_{32}(\tau) = \frac{g_{2}}{g_{1}}\Gamma_{1} ,\\ 
\int_0^{\infty}d\tau e^{\imath\Omega \tau}F_{34}(\tau) = \frac{g_{2}^2}{g_{1}^2}\Gamma_{1},\\\label{eq:Fij_last}
\int_0^{\infty}d\tau e^{\imath\Omega \tau}F_{43}(-\tau) = \frac{g_{2}^2}{g_{1}^2}\Gamma_{2}.
\end{eqnarray}
The remaining terms which are of the form $\mbox{Tr}_B(b_k e^{-\imath \Omega_k t} b_k e^{-\imath \Omega_k s} \rho_B)$ or\\ $\mbox{Tr}_B(b_k^\dagger e^{-\imath \Omega_k t} b_k^\dagger e^{-\imath \Omega_k s }\rho_B)$ are equal to zero since we assumed the bath is in an equilibrium state. After substituting Eq.~(\ref{eq:S_i operators}) and  Eq.~(\ref{eq:Fij}-\ref{eq:Fij_last}) into Eq.~(\ref{eq:twooneone}) and neglecting the contributions from $\delta_1$ and  $\delta_2$, one obtains the master equation, 
\begin{eqnarray}
\label{eq:MASTEREQUATION}
\frac{\partial}{\partial t}\rho_{S}(t) =  -\imath~
[ \frac{\Omega}{2}\sigma_{z} ,\rho_S(t)] +\gamma_{1} \Big(2 \sigma_{-}\rho_{S}(t) \sigma_{+} - \{\sigma_{+}\sigma_{-}, \rho_{S}(t)\}\Big)\nonumber\\ +\gamma_{2} \Big(2 \sigma_{+}\rho_{S}(t) \sigma_{-} - \{\sigma_{-}\sigma_{+}, \rho_{S}(t)\}\Big)\nonumber 
-\imath~[ \Omega a^{\dagger}a,\rho_S(t)]\\ +\gamma_{1}\eta^2 \Big(2a\rho_{S}(t) a^{\dagger} - \{a^{\dagger}a, \rho_{S}(t)\}\Big) +\gamma_{2}\eta^2 \Big(2 a^{\dagger}\rho_{S}(t) a - \{aa^{\dagger}, \rho_{S}(t)\}\Big)\nonumber\\
\left.\begin{array}{r}
+\gamma_{1}\eta \Big(2\sigma_{-}\rho_{S}(t) a^{\dagger} + 2 a\rho_{S}(t) \sigma_{+}- \{\sigma_{+}a, \rho_{S}(t)\} -  \{a^{\dagger}\sigma_{-}, \rho_{S}(t)\}\Big)\\
+\gamma_{2}\eta\Big(2\sigma_{+}\rho_{S}(t) a + 2 a^{\dagger}\rho_{S}(t) \sigma_{-}- \{a\sigma_{+}, \rho_{S}(t)\} -  \{\sigma_{-}a^{\dagger}, \rho_{S}(t)\}\Big)
\end{array}
\right.
\end{eqnarray}
where $\eta = g_{2}/g_{1}$ is the ratio of coupling strengths between qubit-bath coupling and oscillator-bath coupling.

We consider the single excitation sector for the Harmonic Oscillator so as to make use of Positive Partial Transpose criterion (P.P.T) \cite{peres1996separability,horodecki2001separability} in showing that the dynamics described by Eq.~(\ref{eq:MASTEREQUATION}) is entangling. One could consider the second excitation sector of the oscillator which would map the problem to the case of a qubit-qutrit interaction. However, we do not address it in the present study. We denote the eigenstate of lowest energy level by $\ket{0}_{HO}$ and the eigenstate of highest energy level by $\ket{1}_{HO}$ as shown in Figure \ref{fig13}. We replace $a$ and $a^\dagger$ with $\sigma_{-}^{\scaleto{HO}{3pt}}$ and $\sigma_{+}^{\scaleto{HO}{3pt}}$ respectively as they are acting on a qubit. For clarity, we replace  $\sigma_{+}$ and $\sigma_{-}$ with $\sigma_{+}^{\scaleto{Q}{4pt}}$ and $\sigma_{-}^{\scaleto{Q}{4pt}}$ respectively. One then obtains the master equation (\ref{eq:MASTEREQUATION}),
\begin{eqnarray}
\label{eq:MASTEREQUATION2}
\frac{\partial}{\partial t}\rho_{S}(t) = & -\imath~
[ \frac{\Omega}{2}\sigma_{z} ,\rho_S(t)] + L_Q[\rho(t)] -\imath~[ \Omega \sigma_{+}^{\scaleto{HO}{3pt}}\sigma_{-}^{\scaleto{HO}{3pt}},\rho_S(t)] + L_{HO}[\rho(t)] \nonumber \\ &+ L_{QHO}[\rho(t)],
\end{eqnarray}
where 
\begin{eqnarray}
L_Q[\rho(t)] = \gamma_{1} \Big(2 \sigma_{-}^{\scaleto{Q}{4pt}}\rho_{S}(t) \sigma_{+}^{\scaleto{Q}{4pt}} - \{\sigma_{+}^{\scaleto{Q}{4pt}}\sigma_{-}^{\scaleto{Q}{4pt}}, \rho_{S}(t)\}\Big)\nonumber \\+\gamma_{2} \Big(2 \sigma_{+}^{\scaleto{Q}{4pt}}\rho_{S}(t) \sigma_{-}^{\scaleto{Q}{4pt}} - \{\sigma_{-}^{\scaleto{Q}{4pt}}\sigma_{+}^{\scaleto{Q}{4pt}}, \rho_{S}(t)\}\Big),
\end{eqnarray}
\begin{eqnarray}
L_{HO}[\rho(t)] = \gamma_{1}\eta^2 \Big(2\sigma_{-}^{\scaleto{HO}{3pt}}\rho_{S}(t) \sigma_{+}^{\scaleto{HO}{3pt}} - \{\sigma_{+}^{\scaleto{HO}{3pt}}\sigma_{-}^{\scaleto{HO}{3pt}}, \rho_{S}(t)\}\Big)\nonumber\\ +\gamma_{2}\eta^2 \Big(2 \sigma_{+}^{\scaleto{HO}{3pt}}\rho_{S}(t) \sigma_{-}^{\scaleto{HO}{3pt}} - \{\sigma_{-}^{\scaleto{HO}{3pt}}\sigma_{+}^{\scaleto{HO}{3pt}} \rho_{S}(t)\}\Big),
\end{eqnarray}
and
\begin{eqnarray}
&L_{QHO}[\rho(t)]= \gamma_{1}\eta \Big(2\sigma_{-}^{\scaleto{Q}{4pt}}\rho_{S}(t) \sigma_{+}^{\scaleto{HO}{3pt}}  -  \{\sigma_{+}^{\scaleto{HO}{3pt}}\sigma_{-}^{\scaleto{Q}{4pt}}, \rho_{S}(t)\}\Big)\nonumber\\
&+ \gamma_{1}\eta \Big(2 \sigma_{-}^{\scaleto{HO}{3pt}}\rho_{S}(t) \sigma_{+}^{\scaleto{Q}{4pt}}- \{\sigma_{+}^{\scaleto{Q}{4pt}}\sigma_{-}^{\scaleto{HO}{3pt}}, \rho_{S}(t)\} \Big)\nonumber\\
&+\gamma_{2}\eta\Big(2\sigma_{+}^{\scaleto{Q}{4pt}}\rho_{S}(t) \sigma_{-}^{\scaleto{HO}{3pt}} + 2 \sigma_{+}^{\scaleto{HO}{3pt}}\rho_{S}(t) \sigma_{-}^{\scaleto{Q}{4pt}}- \{\sigma_{-}^{\scaleto{HO}{3pt}}\sigma_{+}^{\scaleto{Q}{4pt}}, \rho_{S}(t)\} -  \{\sigma_{-}^{\scaleto{Q}{4pt}}\sigma_{+}^{\scaleto{HO}{3pt}}, \rho_{S}(t)\}\Big)\nonumber.\\
\end{eqnarray}

The terms $ L_Q[\rho(t)]$ and $ L_{HO}[\rho(t)]$ of the master equation (\ref{eq:MASTEREQUATION2}) describe the dissipation of the qubits $Q$ and $HO$ respectively. The term $L_{QHO}[\rho(t)]$ includes the coupling between the qubits $Q$ and $HO$ induced by the bath.

The master equation (\ref{eq:MASTEREQUATION2}) can be decomposed into 16 differential equations, one for each entry of the density matrix. Using this one can derive a dynamical equation of evolution for the density matrix. We use this to look at the dynamics of quantum correlations in the system in Sect.~3 and Sect.~4.
\section{Generation of Quantum Correlations}
\begin{figure*}
	\centering
	\includegraphics[scale = 0.60]{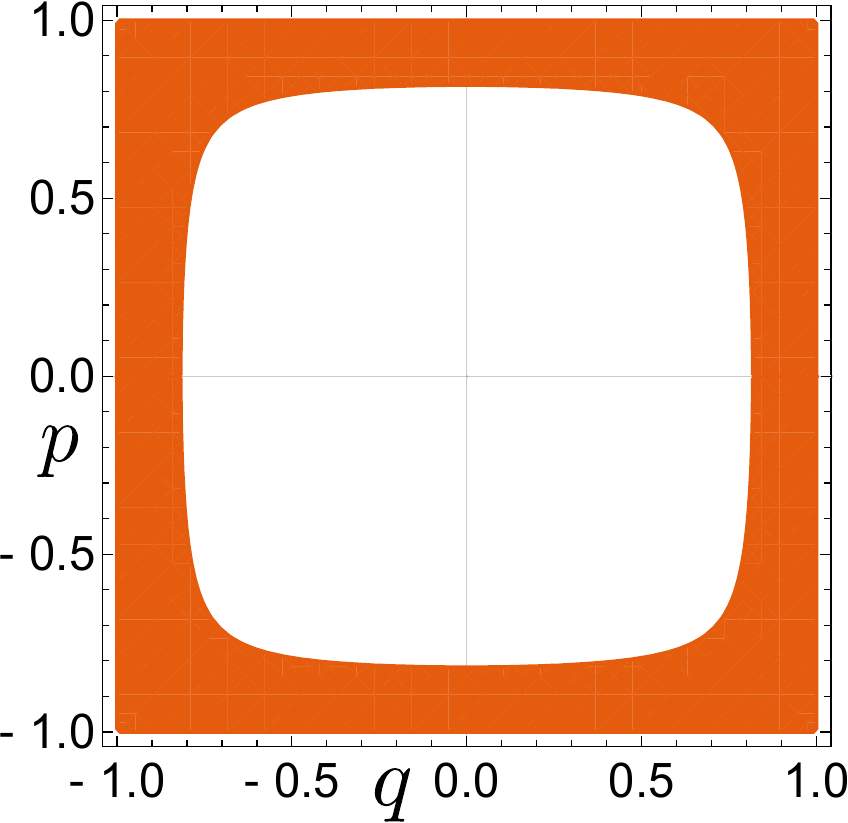}\hspace{5pt}
	\caption{(Color online) The set of all points $(p,q)$, for  which there exists at  least one 3-tuple $(\alpha,\beta,\vartheta )$ such that the condition (\ref{eq:condition2}) is satisfied, are shown above.} \label{fig11}
	
\end{figure*}
Bennati et al. \cite{benatti2008environment,benatti2003environment} give the conditions for a semigroup to entangling. We make use of the result to show that the dynamics described by Eq.~(\ref{eq:MASTEREQUATION2}), which is stemming from a combination of Jaynes-Cummings interaction and dipole interaction as described in Sect.~1, is entangling. We further analytically find out the class of initial states of the system which lead to entanglement as $t \rightarrow 0^+$.

The master equation of the system, which is effectively a pair of two-level systems ($Q$ and $HO$), is in a GKSL \cite{gorini1976completely} form:
\begin{eqnarray}\label{eq:VeryGenMasterEqn}
\partial_t\rho(t) = - \imath[H,\rho(t)]  +L[\rho(t)],
\end{eqnarray}
where $H$, which is equal to $H_S$ in our case, is the effective system Hamiltonian, while
\begin{eqnarray}\label{eq:LindbladForm}
L[\rho(t)] = \sum_{i, j  } \mathcal{K}_{ij}\left[ \mathcal{G}_i \rho(t)  \mathcal{G}_j^\dagger - \frac{1}{2}\{\mathcal{G}_j^\dagger\mathcal{G}_i,\rho(t)\} \right],
\end{eqnarray} 
is the dissipative term  and $\{\mathcal{G}_i \}_{\{i= 1\dots d^2-1\}} $ can be any trace orthonormal basis. By comparing with  Eq.~(\ref{eq:MASTEREQUATION2}), one can see that 
\begin{eqnarray}
&&\mathcal{G}_1 =  \sigma_{+}^{\scaleto{Q}{4pt}}\otimes \mathbbm{1}^{\scaleto{HO}{3pt}},\nonumber\\
&&\mathcal{G}_2 =  \sigma_{-}^{\scaleto{Q}{4pt}}\otimes \mathbbm{1}^{\scaleto{HO}{3pt}},\nonumber\\ 
&&\mathcal{G}_3 =  \mathbbm{1}^{\scaleto{Q}{4pt}}\otimes \sigma_{-}^{\scaleto{HO}{3pt}},\nonumber\\ 
&&\mathcal{G}_4 =  \mathbbm{1}^{\scaleto{Q}{4pt}}\otimes \sigma_{+}^{\scaleto{HO}{3pt}}.
\end{eqnarray}
The co-efficients $\mathcal{K}_{ij}$ form Kossakowski Matrix  ($\mathcal{K}$) \cite{kossakowski72}
\begin{equation}
\mathcal{K} = \mqty(
2\gamma_{2} & 0 & 0 &2\eta\gamma_{2}\\
0& 2\gamma_{1}&2\eta\gamma_{1}&0\\
0& 2\eta\gamma_{1}&2\eta^2\gamma_{1}&0\\ 
2\eta \gamma_{2}&0&0&2 \eta^2\gamma_{2}),
\end{equation}
which is Hermitian i.e. $\mathcal{K} = \mathcal{K}^\dagger$.

The dissipative contribution of $L[\rho(t)]$ is written as
\begin{equation}
L[\rho(t)] = L_Q[\rho(t)] + L_{HO}[\rho(t)] + L_{QHO}[\rho(t)].
\end{equation}
According \cite{hu2011necessary}, the dynamics generate non-classical correlations between the qubits $Q$ and $HO$ since $L_{QHO}[\rho(t)] \neq 0$. This is very intuitive because  $L_{QHO}[\rho(t)]$ includes the induced coupling between the qubits $Q$ and $HO$.

Since non-classical correlations do not imply entanglement, we now specifically check if the dynamics described by Eq.~(\ref{eq:MASTEREQUATION2}) is entangling. According to the Positive Partial Transpose criterion \cite{peres1996separability,horodecki2001separability}, the system $\rho$ is separable if and only if $\rho^{T_{HO}}$ is positive semi definite, where $\rho^{T_{HO}}$ denotes the partial transposition on $\rho$ with respect to the subsystem $HO$. One can then define the following quantity 
\begin{eqnarray}
\label{eq:ExpectationRhoTilde}
\Xi(t) = \bra{\psi}\rho^{T_{HO}}(t)\ket{\psi},
\end{eqnarray}
to look for negative eigenvalues in $\rho^{T_{HO}}(t)$. So $\Xi(t) < 0$ implies $\rho(t)$ is entangled. 

From \cite{benatti2008environment,benatti2003environment}, we know that the evolution described by Eq.~(\ref{eq:MASTEREQUATION2}) generates entanglement if and only if there exist a separable initial state $\rho(0)$ and a vector $\ket{\psi} \in \mathbbm{C}^4$ such that the conditions 
\begin{eqnarray}
\label{eq:condition1}
\Xi(0) = 0&\\
\label{eq:condition2}
\partial_t \Xi(0) < 0&
\end{eqnarray} are satisfied. This also implies that the system starting in the separable state $\rho(0)$ gets entangled as $t \rightarrow 0^+$ because the condition $\partial_t \Xi(0) < 0$ along with $\Xi(0) = 0$ implies that $\lim\limits_{t \rightarrow 0^+} \Xi(t) < 0$. We consider the case where the system starts in $\ket{0}_{Q}\ket{1}_{HO}$ to show that the conditions (\ref{eq:condition1}) and (\ref{eq:condition2}) can be satisfied. We choose $\ket{\psi} = \kappa_1 \ket{0}_{Q}\ket{0}_{HO} + \kappa_2 \ket{1}_{Q}\ket{0}_{HO} + \kappa_3\ket{1}_{Q}\ket{1}_{HO}$ according to condition Eq.~(\ref{eq:condition1}).
For this $\ket{\psi}$,
\begin{equation}
\partial_t \Xi(0) = 2 \gamma_{1}\kappa_1^2 \eta^2 - 2(\gamma_{1}+\gamma_{2})\kappa_1\kappa_3\eta +  2\gamma_{2}\kappa_3^2.
\end{equation}
The above equation, which is quadratic in $\kappa_1$(or $\kappa_3$) has two roots unless $\gamma_{1} = \gamma_{2}$.
Since one can find $\kappa_1$ and $\kappa_3$ for a given $\gamma_{1},\gamma_{2}$ and $\eta$ such that condition (\ref{eq:condition2}) is satisfied, the system gets entangled and the considered dynamics in Eq.~(\ref{eq:MASTEREQUATION2}) are entangling. For example, when  $\gamma_1 = 1.01, \gamma_2 = 0.01$ and $\Omega = 0.001$, one can see that  $\kappa_1 = 1$ and  $\kappa_3 = - 1/2$ satisfy condition (\ref{eq:condition2}).

We consider the following general form for the initial pure and separable state of the system
\begin{eqnarray}\label{eq:generalinitialstate}
\ket{\phi} = (p\ket{0} + \sqrt{1-p^2}\ket{1})_{Q} \otimes(q\ket{0} + \sqrt{1-q^2}\ket{1})_{HO},
\end{eqnarray}
where $p,q \in [ -1,1]$,  to find the subset of states which generate entanglement as $t \rightarrow 0^+$ when evolved according to Eq.~(\ref{eq:MASTEREQUATION}). We first choose a  general $\ket{\psi}$ which satisfies condition (\ref{eq:condition1})
\begin{align}
\label{eq: GeneralFormofPsiforInitialstate}
\ket{\psi} = \alpha \left( ( \sqrt{1-p^2}\ket{0} - p\ket{1} ) \otimes(q\ket{0} + \sqrt{1-q^2}\ket{1}) \right)\nonumber\nonumber\\  
+\beta \left((p\ket{0} + \sqrt{1-p^2}\ket{1})   \otimes( \sqrt{1-q^2}\ket{0} - q\ket{1} )\right) \nonumber\\ 
+ \vartheta\left(( \sqrt{1-p^2}\ket{0} - p\ket{1} ) \otimes ( \sqrt{1-q^2}\ket{0} - q\ket{1} ) \right).
\end{align}
For this $\ket{\psi}$  and $\rho(0) = \ketbra{\phi}$, we find
\begin{eqnarray}
\label{eq:generalinequalityofPandQ} 
\begin{array}{c}
\partial_t \Xi(0) = 2 \left(\gamma _2 p^4+\left(p^2-1\right)^2 \gamma _1\right) \alpha ^2 +2 \beta ^2 \eta ^2 \left(\gamma _2 q^4+\left(q^2-1\right)^2 \gamma _1\right)\\ -2 \left( \left(p^2 \left(2 q^2-1\right)-q^2\right) \beta  \eta  \left(\gamma _1+\gamma _2\right) \alpha\right). \\
\end{array}
\end{eqnarray}
For the initial settings $\gamma_1 = 1.01, \gamma_2 = 0.01$ and $\Omega = 0.001$, the Eq.~(\ref{eq:generalinequalityofPandQ}) is reduced to 
\begin{eqnarray}
\begin{array}{c}
\partial_t \Xi(0) = \frac{1}{50}\left( \alpha ^2 \left(102p^4-202 p^2+101\right) + 102 \alpha  \beta  \eta  \left(p^2 \left(1-2 q^2\right)+q^2\right)\right)\\
+\frac{1}{50}\left(\beta ^2 \eta ^2 \left(102 q^4-202q^2+101\right)\right).\\
\end{array}
\end{eqnarray}
One can find the set of all $(p,q)$, where $p,q \in [-1,1]$, for which there exists at  least one 3-tuple $(\alpha,\beta,\vartheta )$ such that the condition (\ref{eq:condition2}) is satisfied. These set of points $(p,q)$ are plotted in Figure \ref{fig11}.

Note that even if one includes the contributions of $\delta_1$ and $\delta_2$, which are the bath-induced Lamb shifts, the above result is not affected due the fact that the quantity $\partial_t \Xi(0)$ (Eq.~(\ref{eq:generalinequalityofPandQ})) remains the same. 
\section{Dynamics of the correlations}
\begin{figure*}
	\centering
	\begin{minipage}{0.5\textwidth}
		\includegraphics[scale = 0.31]{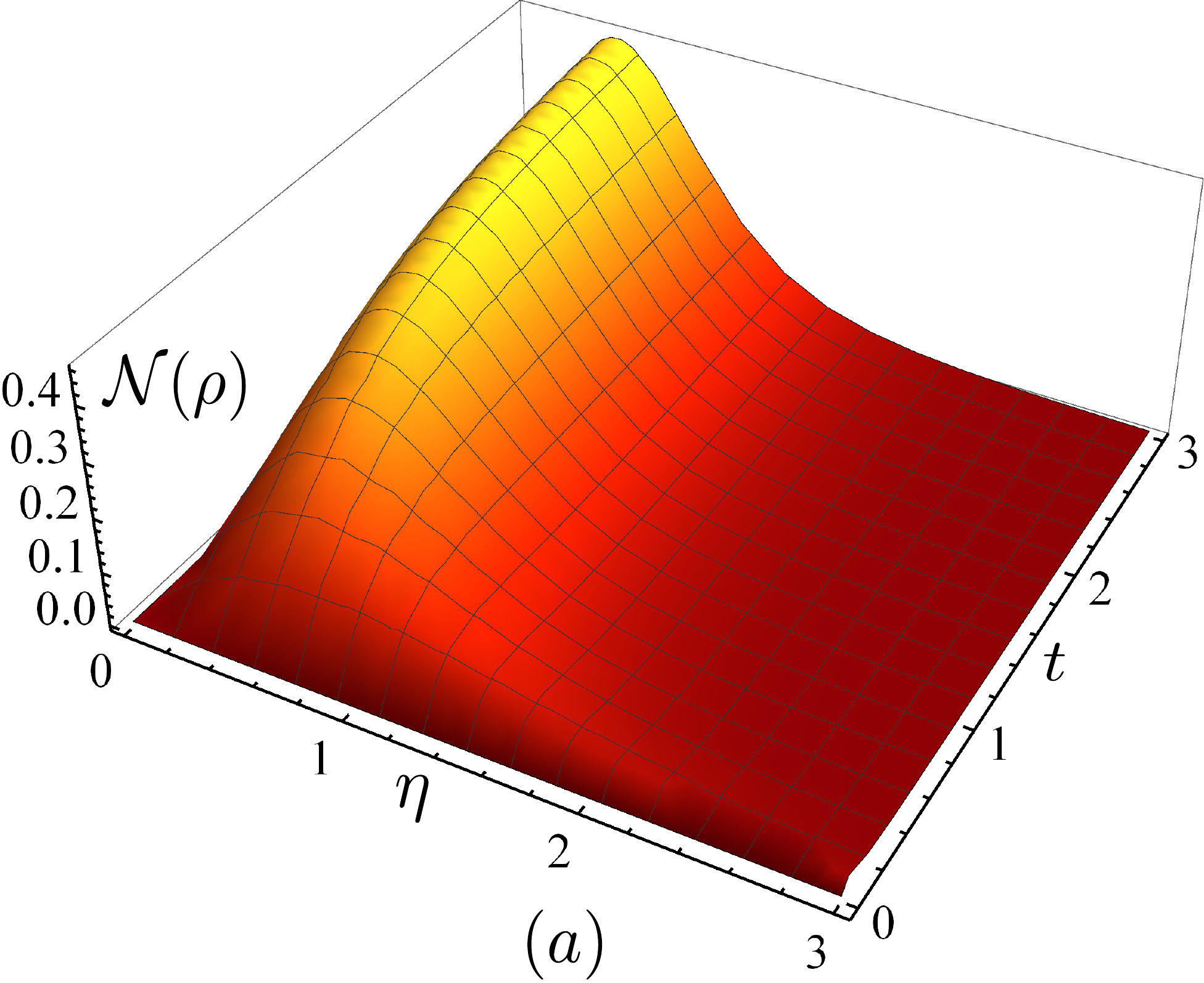}
	\end{minipage}\hfill
	\begin{minipage}{0.5\textwidth}
		\includegraphics[scale = 0.31]{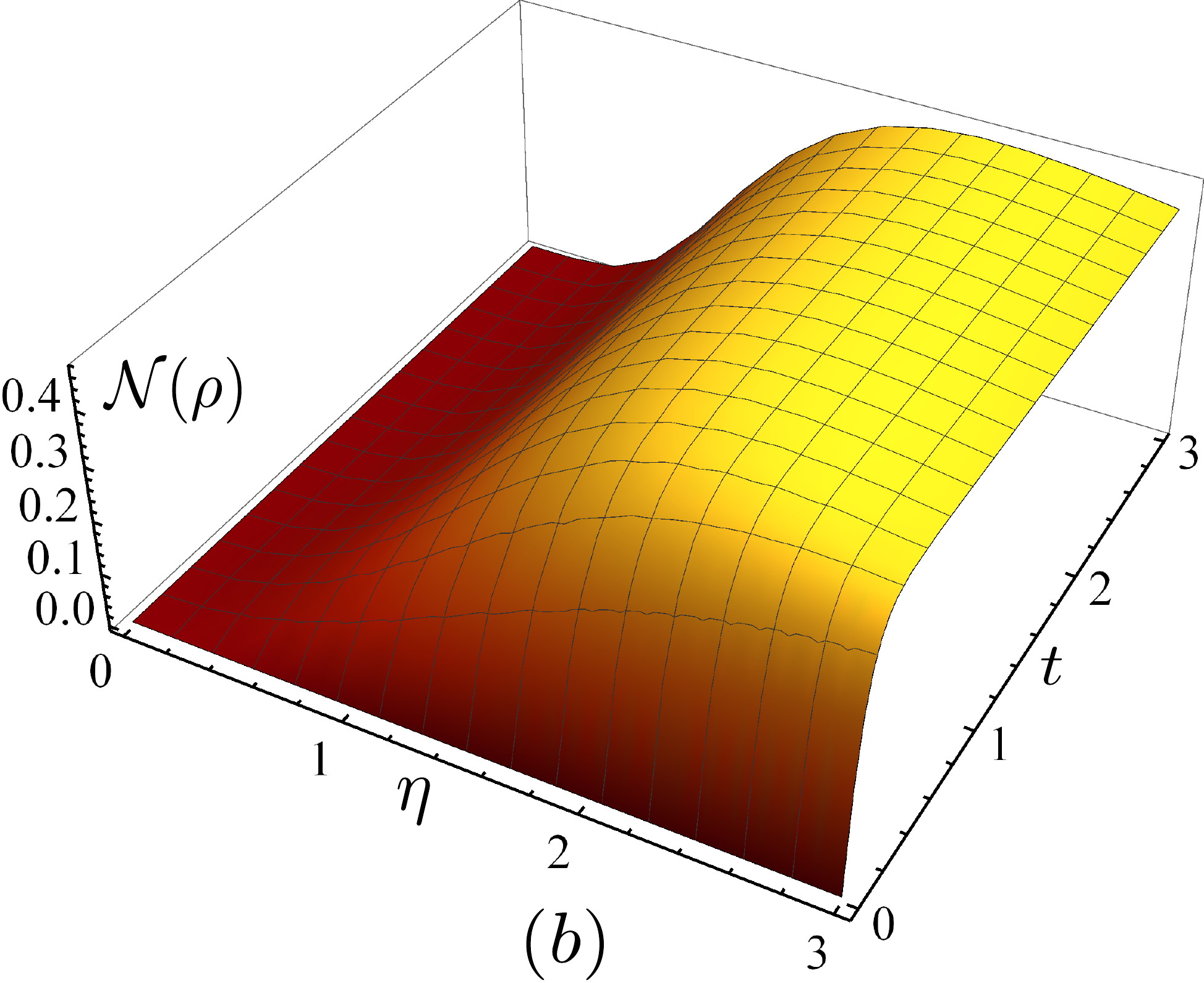}
	\end{minipage}\hfill
	\caption{(Color online) Negativity $\mathcal{N(\rho)}$ of the system, consisting of two qubits ($Q$ and $HO$), as a function of time $t$ and ratio of coupling strengths between qubit-bath coupling and oscillator-bath coupling $\eta$ with parameters $\Omega$ = 0.001, $\gamma_{1} = 1.01$ and $\gamma_{2} = 0.01$. (a) refers to the system starting in $\ket{0}_{Q}\ket{1}_{HO}$. (b) refers to the system starting in $\ket{1}_{Q}\ket{0}_{HO}$.} 
	\label{fig1}
\end{figure*}
In this section, the dynamics of quantum correlations in the two effective qubit system interacting with the bath are studied with respect to $\zeta t$ which we denote by $t$ for brevity.
\begin{figure*}
	
	\begin{minipage}{0.47\textwidth}
		\includegraphics[scale = 0.30]{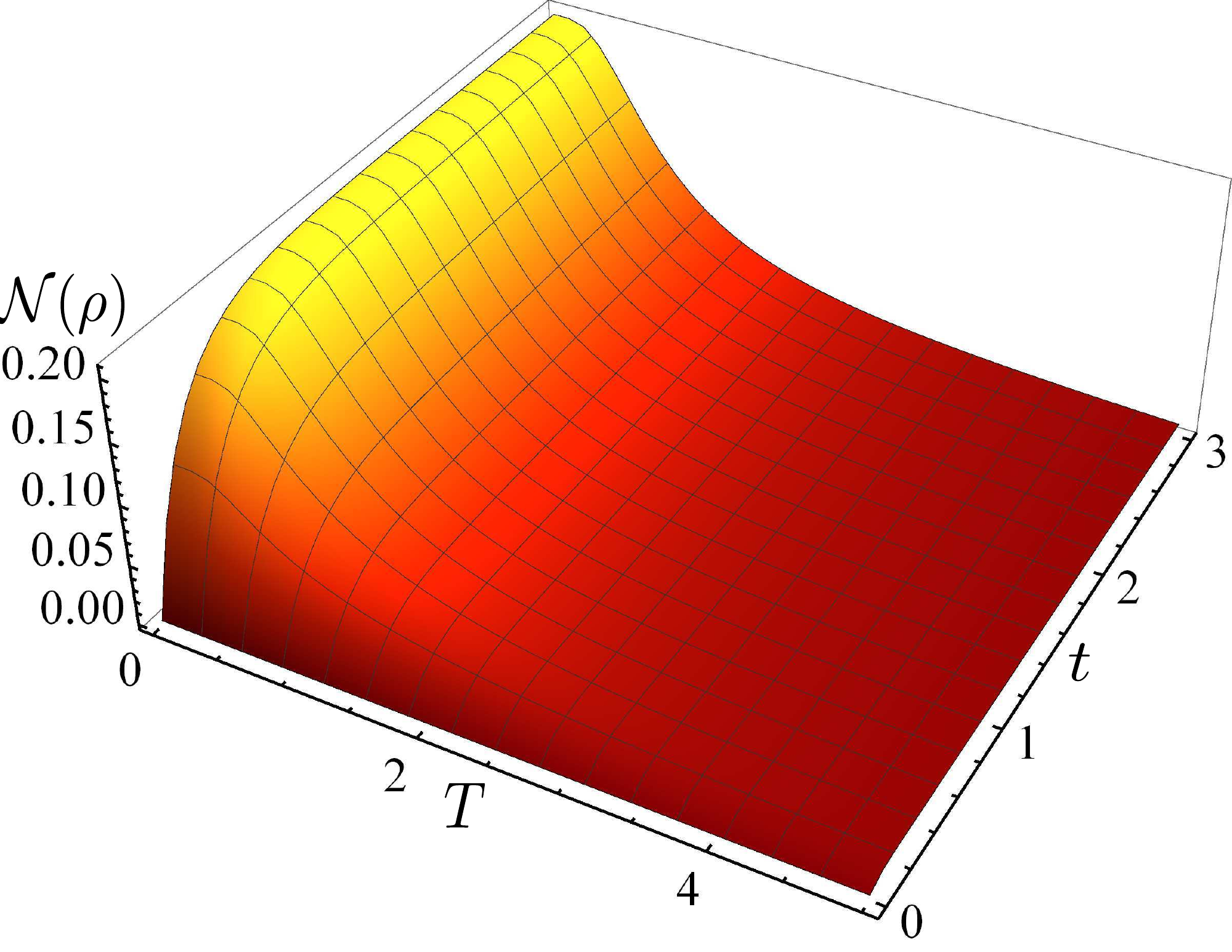}\hspace{5pt}
		\caption{(Color online) Negativity $\mathcal{N(\rho)}$ of the system as a function of time $t$  and temperature of the bath $T$. The system is starting in $\ket{0}_{Q}\ket{1}_{HO}$ state and the parameters $\Omega = 0.001$. We assume $\eta = 1$ for simplicity.} \label{fig7}
	\end{minipage}\hfill
	\begin{minipage}{0.47\textwidth}
		\includegraphics[scale = 0.53]{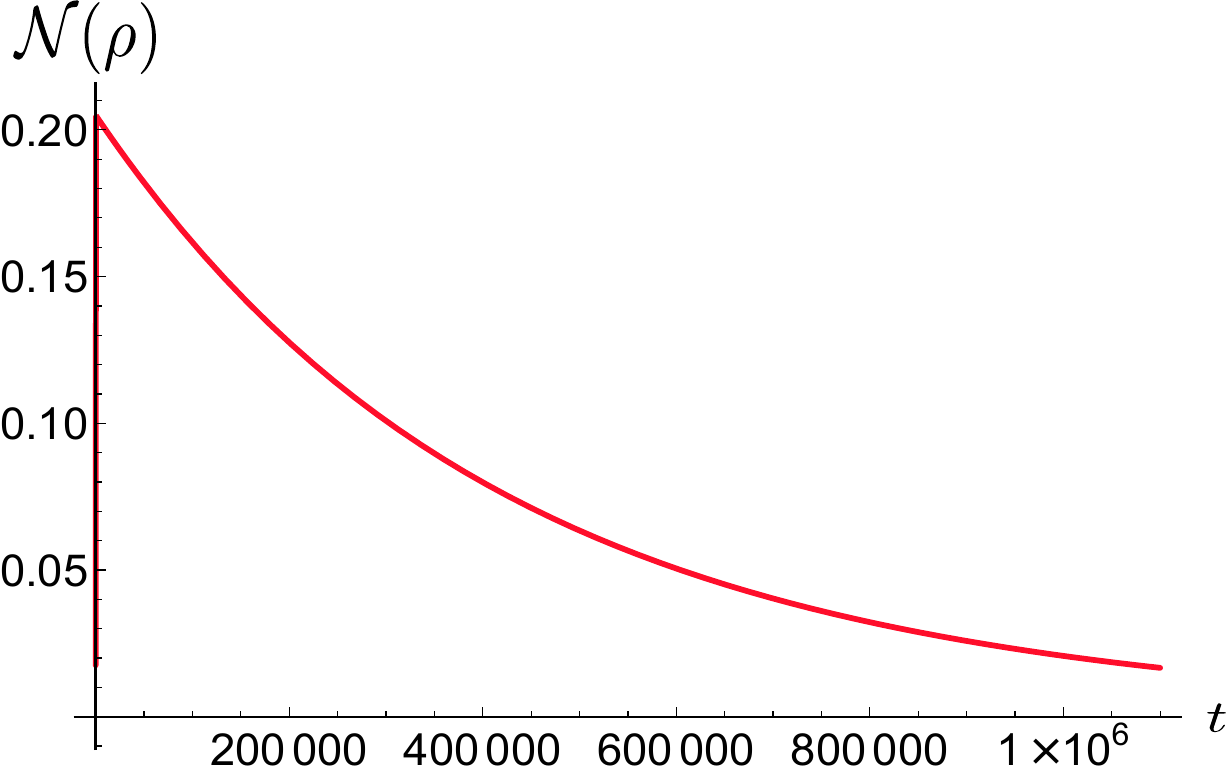}\hspace{5pt}
		\caption{(Color online) Negativity $\mathcal{N(\rho)}$ of the system, starting in  $\ket{0}_{Q}\ket{1}_{HO}$ state with parameters   $\Omega$ = 0.001, $\gamma_{1} = 1.01$ and $\gamma_{2} = 0.01$, over a large period of time $t$. We assume $\eta = 1$ for simplicity. \label{fig5} } 
	\end{minipage}
	
\end{figure*}

\subsection{Entanglement}	
For a state $\rho$, which has two subsystems $A$ and $B$, Negativity \cite{eisert2006entanglement,vidal2002computable}  is defined as
\begin{eqnarray}
\label{eq:Negativity_1}
\mathcal{N}(\rho) := \frac{\norm{\rho^{T_A}}_1 - 1}{2},
\end{eqnarray}
where $\rho^{T_A}$ is the partial transpose of $\rho$ with respect to the subsystem $A$ and $\norm{X}_1$ is the trace norm or sum of the singular values of any arbitrary operator $X$. Alternatively, Negativity is defined as sum of negative eigenvalues of $\rho^{T_A}$ 
\begin{eqnarray}
\label{eq:Negativity_2}
\mathcal{N}(\rho) = \sum_i\frac{\abs{\lambda_i} -\lambda_i }{2},
\end{eqnarray}
where $\lambda_i$ are the eigenvalues. Note that the above definition is equivalent to the first definition Eq.~(\ref{eq:Negativity_1}).

In Figure \ref{fig1}, we plot the Negativity of the system, starting in an initial state $\ket{0}_{Q}\ket{1}_{HO}$ in Figure \ref{fig1}(a) and  $\ket{1}_{Q}\ket{0}_{HO}$ in Figure \ref{fig1}(b), with respect to time and relative coupling strengths of qubit-bath coupling and oscillator bath coupling ($\eta$) for parameters $\Omega = 0.001 , \gamma_{1} = 1.01$ and $\gamma_{2} = 0.01$.  We see that for $\eta> 0$ the entanglement of the system builds up with time as a result of the coupling mediated by the bath which is in complete agreement with our result in Sect.~3 since the states  $\ket{0}_{Q}\ket{1}_{HO}$ and  $\ket{1}_{Q}\ket{0}_{HO}$ are present in the region showed in Figure \ref{fig11}.

\begin{figure*}
	\centering
	\includegraphics[scale = 0.3]{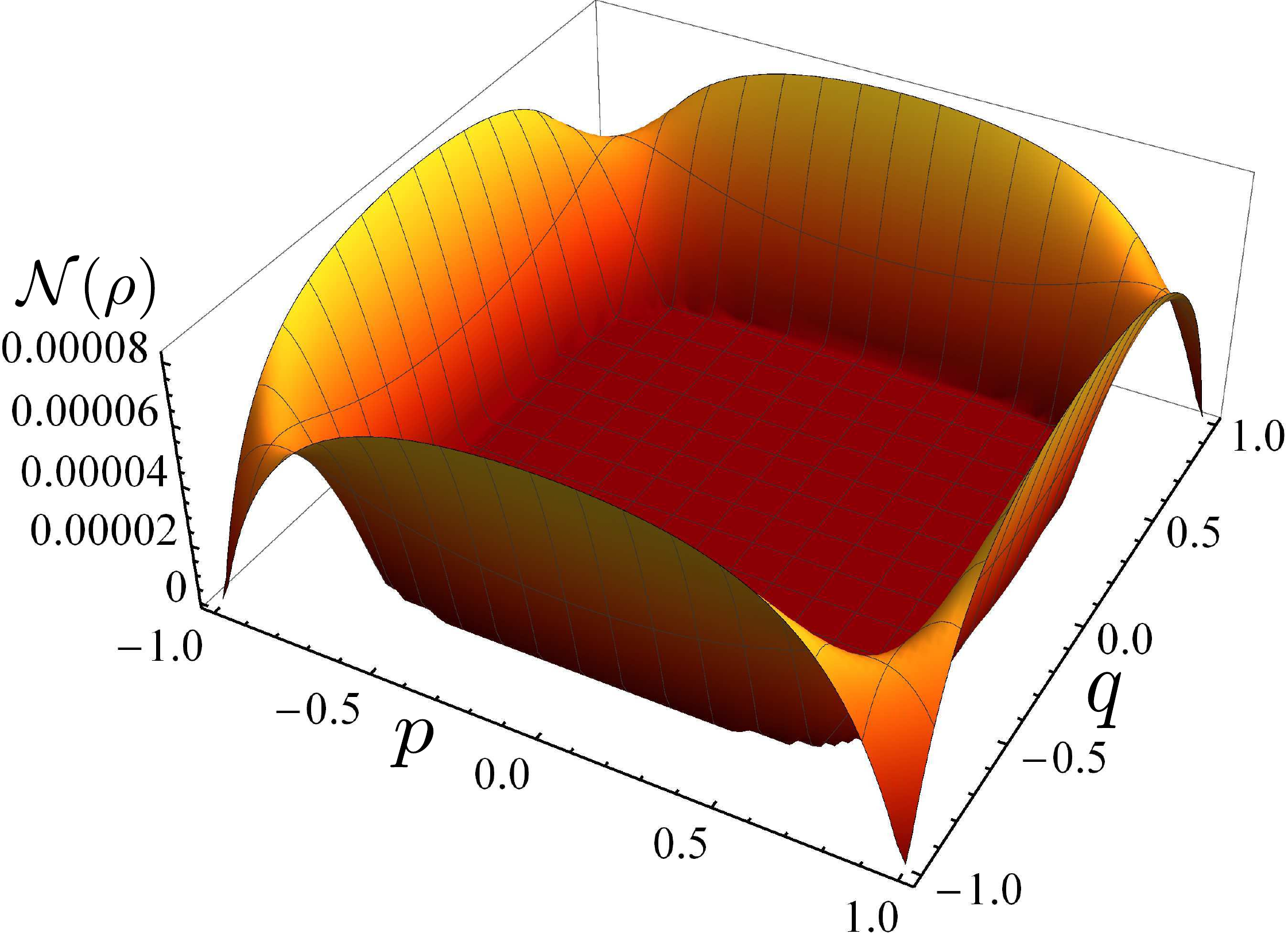}\hspace{5pt}
	\caption{(Color online) Negativity $\mathcal{N(\rho)}$ of the system at $t = 0.0001$ starting in the initial state of the form (\ref{eq:generalinitialstate}) with parameters $\Omega$ = 0.001, $\gamma_{1} = 1.01$ and $\gamma_{2} = 0.01$. We assume $\eta = 1$ for simplicity. }  \label{fig12}
\end{figure*}

In Figure \ref{fig12}, we plot the Negativity  $\mathcal{N(\rho)}$ of the system at time $t = 0.0001$, for all possible initial states of the  form Eq.~(\ref{eq:generalinitialstate}). We see that projection of Negativity  $\mathcal{N(\rho)}$ on to $pq$ plane is exactly same as Figure \ref{fig11}.

In Figure \ref{fig7}, we show how the entanglement between the two qubits varies with the temperature of the bath $T$. We see that entanglement generation decreases as the temperature of the bath increases. Hence the initial choice of bath's temperature is crucial for entanglement generation. One can see that the entanglement persists for a long time  $t$ and goes to zero (Figure \ref{fig5}) as the master equation (\ref{eq:MASTEREQUATION}) has a unique steady state $\rho_{SS}$,
\begin{equation}
\rho_{SS} = \frac{1}{ (1 + \frac{\gamma_{1}}{\gamma_{2}})^2}\mqty((\frac{\gamma_{1}}{\gamma_{2}})^2& 0& 0& 0\\
0 & \frac{\gamma_{1}}{\gamma_{2}}& 0 &0 \\
0&0&\frac{\gamma_{1}}{\gamma_{2}}&0\\
0&0&0&1),	 
\end{equation}
which is clearly separable.

\subsection{Mutual Information}
Mutual Information $I(A:B)$ quantifies the total correlations between the two subsystems of a bipartite system \cite{henderson2001classical}. It is defined as
\begin{eqnarray}
I(A:B) = S(\rho_A) + S(\rho_B) - S(\rho_{AB}),
\end{eqnarray}
where $S(\rho) = -\mbox{Tr}(\rho\log(\rho))$ is the Von-Neumann entropy. 
In Figure \ref{fig9}, we look at the dynamics of Mutual Information $I(Q:HO)$ of system, with respect to time and ratio of coupling strengths between qubit-bath coupling and oscillator-bath coupling $\eta$,  when it is starting in $\ket{0}_{Q}\ket{1}_{HO}$ and $\ket{1}_{Q}\ket{0}_{HO}$. By comparing  Figure \ref{fig9} to Figure \ref{fig1}, one can see that Mutual Information attains higher values at almost every instant of time compared to Negativity as it quantifies both quantum and classical correlations.
\subsection{Quantum Discord}
One can also look at the dynamics of correlations which are weaker than entanglement like quantum discord. Quantum Discord $\delta(AB:B)$ \cite{henderson2001classical,ollivier2001quantum} quantifies the difference between total correlations ($I(A:B)$) and classical correlations ($\max_{\Pi_i^B} J(\rho^{AB})_{\{\Pi_i^B\}}$) in the system. The quantity $ J(\rho^{AB})_{\{\Pi_i^B\}}$ is defined as 
\begin{eqnarray}
J(\rho^{AB})_{\{\Pi_i^B\}} = S(\rho_A) - S(\rho_A|\{\Pi_i^B\}),
\end{eqnarray}
\begin{figure*}
	\begin{minipage}{0.5\textwidth}
		\includegraphics[scale = 0.31]{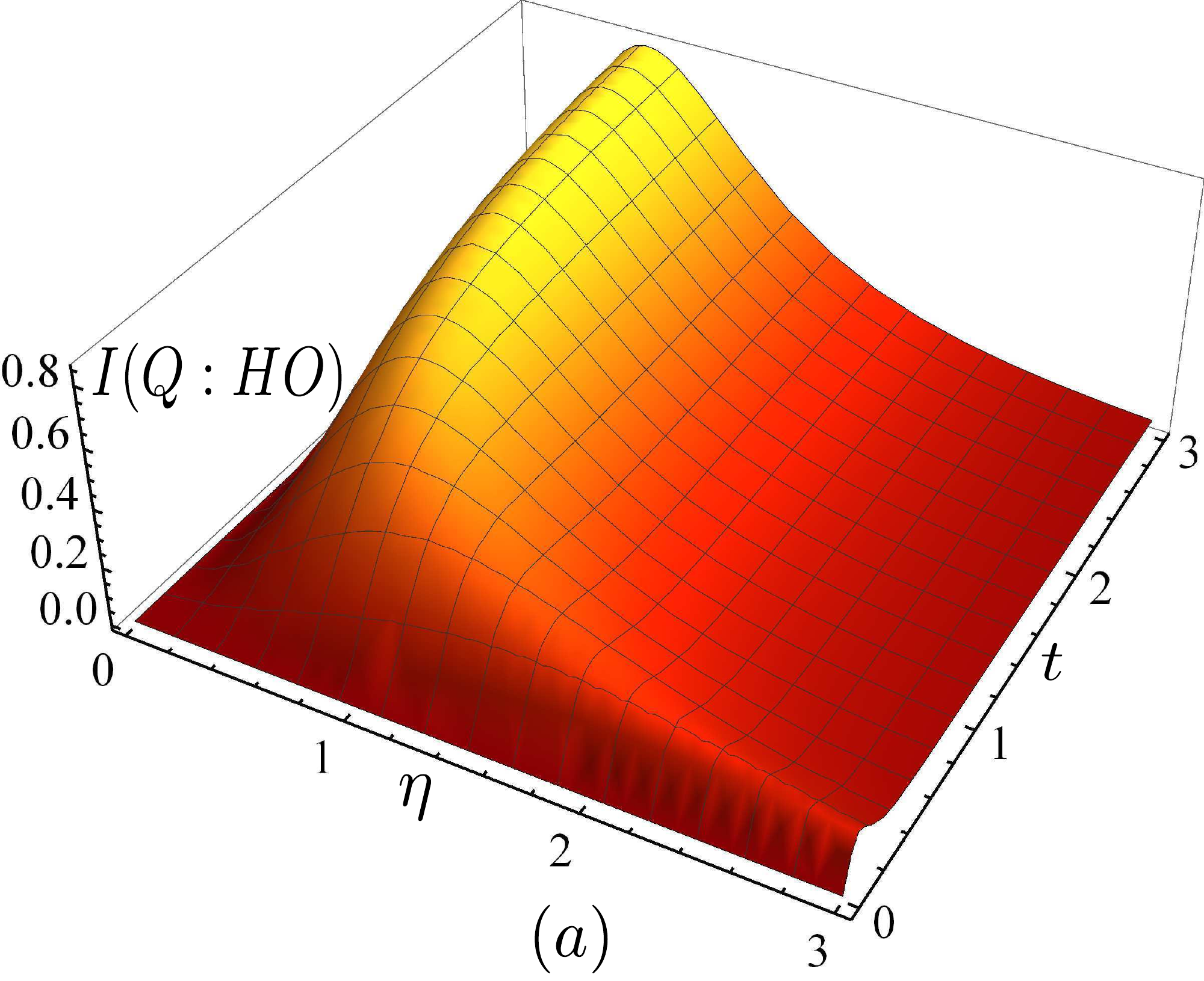}
	\end{minipage}\hfill
	\begin{minipage}{0.5\textwidth}
		\includegraphics[scale = 0.31]{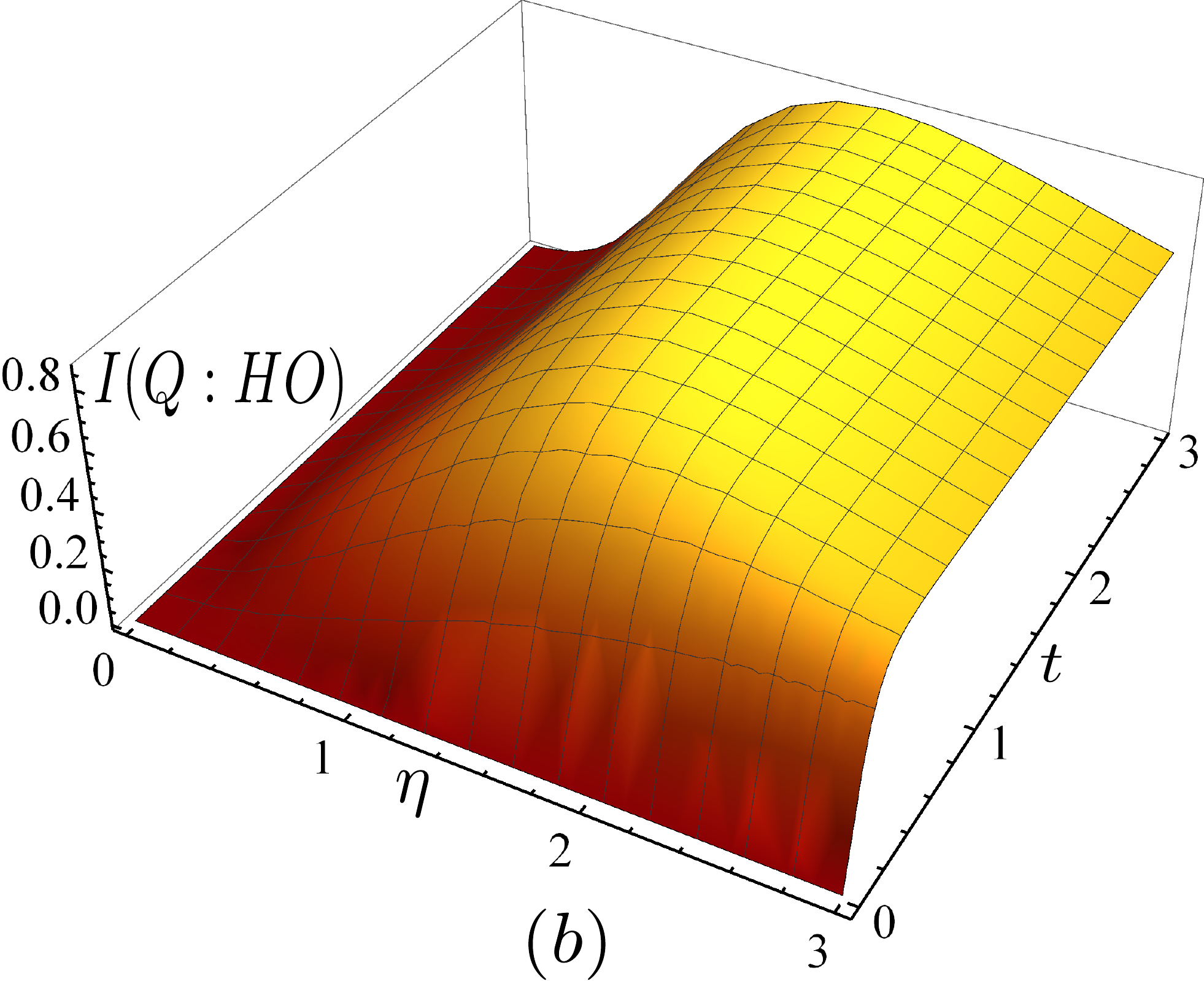}
	\end{minipage}\hfill
	\caption{(Color online) Mutual Information $I(Q:HO)$ of the system, consisting of two qubits ($Q$ and $HO$), as a function of time $t$ and ratio of coupling strengths between qubit-bath coupling and oscillator-bath coupling $\eta$ with parameters $\Omega$ = 0.001, $\gamma_{1} = 1.01$ and $\gamma_{2} = 0.01$. (a) refers to the system starting in $\ket{0}_{Q}\ket{1}_{HO}$. (b) refers to the system starting in $\ket{1}_{Q}\ket{0}_{HO}$. } \label{fig9} 
\end{figure*}
where $ S(\rho_A|\{\Pi_i^B\})$ is the conditional entropy of $A$, with respect to the on-Neumann measurements  $\{\Pi_i^B\}$ on $\rho_B$. Quantum Discord  
$\delta(AB:B)$ is then defined as follows,
\begin{eqnarray}\label{eq:qdiscord1}
\delta(AB:B) = \min_{\Pi_i^B}\left[[I(A:B) - J(\rho^{AB})_{\{\Pi_i^B\}}\right],
\end{eqnarray}
where $\rho^{AB}$ is the density matrix of composite system $AB$.
In Figure \ref{fig10}, we look at dynamics of discord of the system, with respect to second subsystem $HO$, when it is starting in $\ket{0}_{Q}\ket{1}_{HO}$ and $\ket{1}_{Q}\ket{0}_{HO}$. For simplicity, we assume that the coupling constants are equal i.e. $g_{1} = g_{2}$. One can see that discord $\delta(S:HO)$ increases with time as a result of quantum correlations being induced by the common bath between the subsystems. 
The set of initial states which generate quantum discord as $t\rightarrow 0^+$ when evolved according to Eq.~(\ref{eq:MASTEREQUATION2}) include the set of states in Figure (\ref{fig11}) and more. This is because of the fact that non-zero entanglement implies non-zero quantum discord but not the vice versa.
One can see from Figure \ref{fig1}, Figure \ref{fig9} and Figure \ref{fig10} that they only differ in the scale. Mutual Information attains the highest values among the three at almost every instant of time as it quantifies both classical and quantum correlations. It is followed by quantum discord which quantifies only quantum correlations. It is further followed by Negativity as it quantifies only stronger quantum correlations named entanglement.
\begin{figure*}
	\centering
	\begin{minipage}{0.50\textwidth}
		\includegraphics[scale = 0.32]{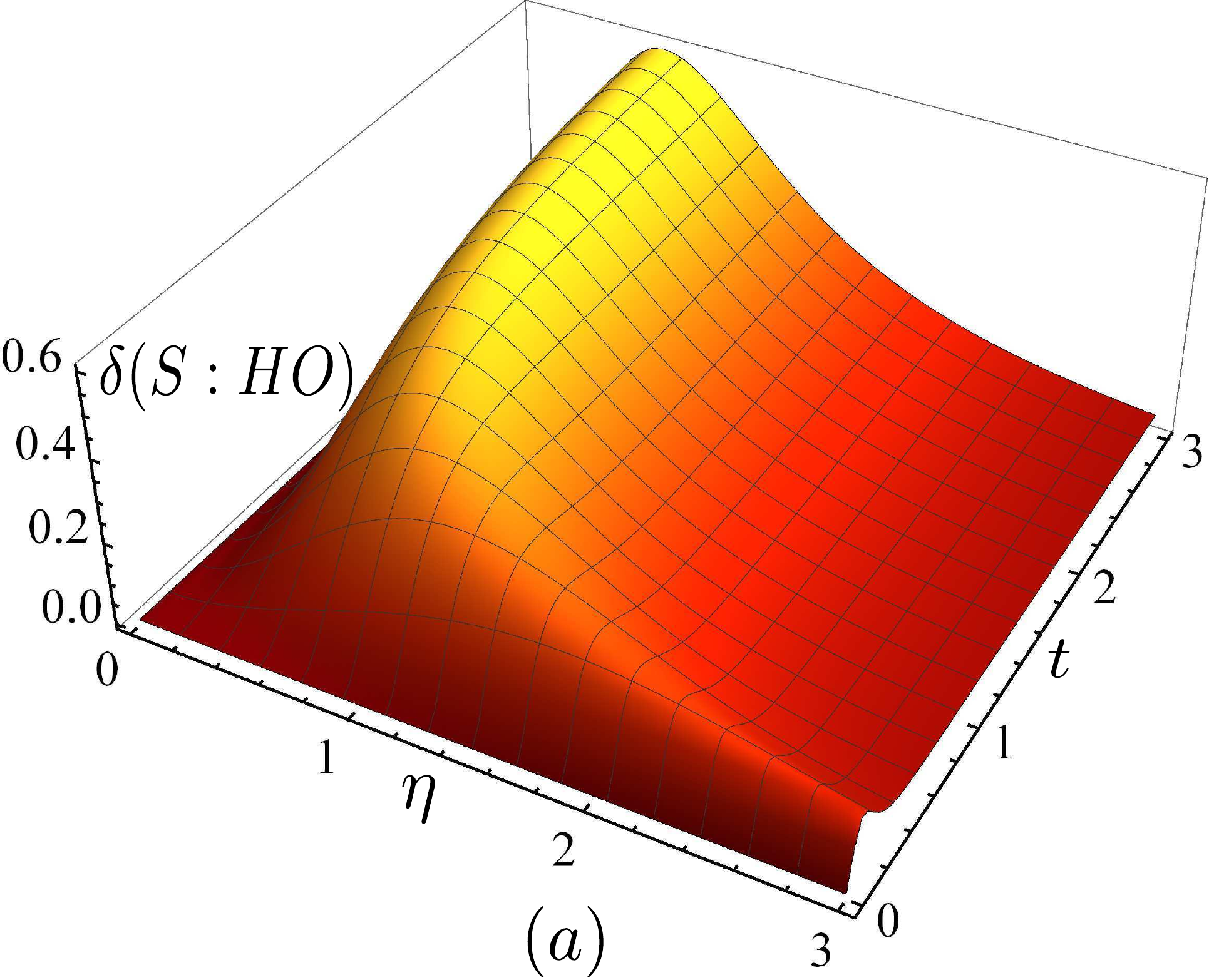}
	\end{minipage}\hfill
	\begin{minipage}{0.50\textwidth}
		\includegraphics[scale = 0.32]{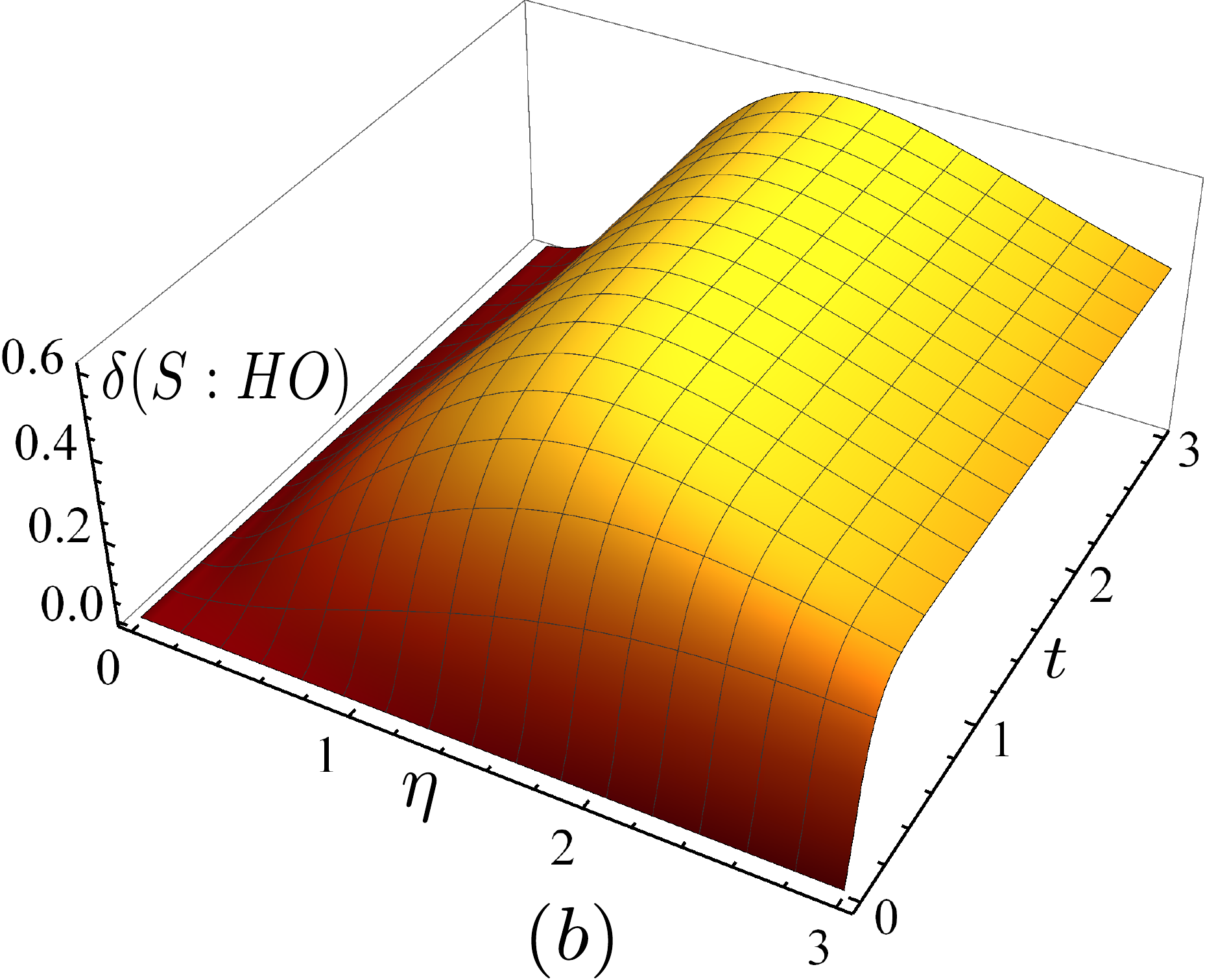}
	\end{minipage}\hfill
	\caption{(Color online) Quantum Discord $\delta(S:HO)$ of the system, consisting of two qubits ($Q$ and $HO$), as a function of time $t$ and ratio of coupling strengths between qubit-bath coupling and oscillator-bath coupling $\eta$ with parameters $\Omega = 0.001$, $\gamma_{1} = 1.01$ and $\gamma_{2} = 0.01$. (a) refers to the system starting in $\ket{0}_{Q}\ket{1}_{HO}$. (b) refers to the system starting in $\ket{1}_{Q}\ket{0}_{HO}$. } \label{fig10}
\end{figure*}
\section{Conclusion}
We study the quantum correlations between a qubit and a harmonic oscillator, not interacting directly, but coupled to a common bath. To this end, we first obtain the microscopic derivation of the master equation for the open dynamics of the system, modelled as a qubit and a harmonic oscillator, with a single excitation sector, interacting with the bath via Jaynes-Cummings interaction type. We make use of the necessary and sufficient conditions given in \cite{hu2011necessary,benatti2008environment,benatti2003environment} to show these dynamics generate non-classical correlations including entanglement. We further analytically find out the class of initial states which generate entanglement. One can do a similar study by considering the second excitation sector for the harmonic oscillator which effectively makes the system a composition of a qubit and qutrit. Entanglement is quantified using negativity \cite{eisert2006entanglement,vidal2002computable} and its dependence on the temperature of the bath is shown. Other measures of quantum correlations studied include mutual information and quantum discord. We note that the entanglement generated persists for long time, making the system potentially useful for practical applications.
\section{Acknowledgments}  
The authors would like to thank Dr. Camille Lombard Latune for insightful comments and suggestions. The work of V.~J.   and F.~P.  is based upon research
supported  by  the South  African  Research  Chair Initiative  of  the
Department of Science and Technology and National Research Foundation. R.~S.  thanks DST-SERB, Govt.  of India,
for financial support provided through the project
EMR/2016/004019.

\end{document}